\documentclass[12pt,preprint,a4paper]{revtex4-1}
\usepackage{amsmath,amssymb,graphicx}
\begin{document}
\title{Simulations of electromagnetic effects in
high frequency capacitively coupled discharges
using the Darwin approximation}
\author{Denis Eremin, Torben Hemke, Ralf Peter Brinkmann, 
and Thomas Mussenbrock}
\affiliation{Ruhr University Bochum, Department of
Electrical Engineering and Information Sciences, 
Institute of Theoretical Electrical Engineering, 
D-44780 Bochum, Germany}
\begin{abstract}
The Darwin approximation is investigated for its
possible use in simulation of electromagnetic effects in large size, high
frequency capacitively coupled discharges.
The approximation is utilized within the framework of
two different fluid models which are applied to
typical cases showing pronounced standing wave and skin effects. 
With the first model it is demonstrated
that Darwin approximation
is valid for treatment of such effects in the range 
of parameters under consideration. The second approach, a reduced
nonlinear Darwin approximation-based model, shows 
that the electromagnetic phenomena persist in a
more realistic setting. The Darwin approximation offers
a simple and efficient way of carrying out electromagnetic
simulations as it removes the Courant condition 
plaguing explicit electromagnetic algorithms and
can be implemented as a straightforward modification
of electrostatic algorithms. The algorithm described 
here avoids iterative schemes needed for the divergence
cleaning and represents a fast and efficient solver,
which can be used in fluid and kinetic models for
self-consistent description of technical plasmas
exhibiting certain electromagnetic activity.
\end{abstract}
\maketitle


\section{Introduction}


Low temperature plasmas play a crucial role in materials
processing \cite{lieberman2005}. Particularly capacitive
radio frequency discharges are important.
Such plasmas, which were first introduced in the 1960s,
have undergone a continuous development keeping up with 
the increasing requirements of the hitech industry.
One trend is the attempt to increase productivity by 
increasing the size of the processed wafers or substrates. 
In microelectronics, wafer sizes of $300$ mm are now state 
of the art, the transition to $450$ mm is envisioned \cite{samukawa2012}. The 
manufacturing of flat panel displays or photovoltaic solar cells 
requires discharges in the meter range \cite{park2003, perrin2000, 
schmitt2002}. The size of the processing equipment obviously
must meet these requirements. A second recent trend is the use of 
higher, often multiple frequencies of up to few hundred MHz
believed to ensure better power coupling, higher plasma density,
and thus higher productivity as well \cite{goto1992, makabe1995}. 

These developments have implications on the physical 
models which can be used for realistic simulations 
of plasma-based manufacturing processes. One such implication 
is of particular importance: The simultaneous increase of 
driving frequency, plasma density, and discharge size puts 
the plasmas into a regime where the commonly used ``electrostatic 
approximation'' of Maxwell's equations no longer suffices.
That means that $\nabla\times\vec E=0$ (and thus $\vec E = 
- \nabla\Phi$) is not justified anymore. (In this
regime the skin depth and/or wavelength of the surface 
waves propagating in the sheath regions may become comparable 
or smaller than the reactor dimension). Recent experimental
studies have revealed that strong plasma and field
non-uniformities appear to be connected with electromagnetic 
effects \cite{sansonnens2003, howling2004, perret2003,
chabert2004-1, hebner2006, miller2006}.

The theoretical aspects of electromagnetic effects
in large-area, high frequency discharges were first
systematically studied by Lieberman et al. \cite{lieberman2002}.
They considered an axisymmetric slab model with uniform
plasma density in the bulk and the matrix sheath (i.e., 
constant ion density and zero electron density), treating
the plasma as a dielectric and assuming the sheath width
to be constant in time. Also, a ``natural boundary"
condition for a CCP discharge with the dielectric sidewall
that all the RF current flows through the discharge (so 
that the magnetic field at the sidewall is constant in 
the axial direction and no external source of electromagnetic
excitation is involved) was employed.
Under these conditions,
they treated the problem analytically 
and gave criteria, based on the estimate of change in
the power deposition profile produced by Ohmic heating
due to these effects, for significance of the standing wave
and skin effects. In essence, a standing wave
consists of surface waves, see, e.g., \cite{bowers2002},
excited in the cavities formed in the powered (grounded)
sheath, bounded  by the electrode, powered (grounded)
sheath interface and the sidewall; the surface
waves in the grounded and powered regions are
weakly coupled with each other through the bulk.
The kind of skin effect meant here manifests itself in tendency
of the induced electric field in an RF-driven discharge to reduce 
current at the axis of the discharge and increase it close
to the discharge periphery  \cite{mussenbrock2008},
similar to the skin effect in metals conducting RF current 
(see, e.g., \cite{ramo1994}).

Several detailed studies were conducted by Chabert et al. 
\cite{chabert2004-2} who used a self-consistent transmission
line model to account for the interaction of RF 
heating and  plasma density. In recent years the
number of contributions to the field has multiplied, for a
review see, e.g., Ref. \cite{chabert2007}. 

The analytical models mentioned above, being significantly
limited by the underlying assumptions, leave many important
questions unanswered. Numerical simulations should be used
to describe the phenomena in more realistic situations.   
So far a number of numerical studies based on the fluid
models have been conducted for the large scale CCP discharges
driven at high frequencies \cite{chen2010, rauf2008, yang2010, 
lee2008, zhang2010}. All those approaches were based
on the full set of Maxwell's equations, which contains
electromagnetic radiation. The underlying equations are 
hyperbolic and one has to satisfy the Courant criterion
on numerical stability provided an explicit algorithm,
such as the popular finite-difference time-domain (FDTD) algorithm,
is used for their numerical solution. The Courant criterion
forces the time step to be small and the overall simulation
time very long. One can resort to implicit algorithms for the
numerical solution of Maxwell's equations, however, it
must be noted that the existing implicit numerical schemes
are rather intricate. 

In this paper we propose an alternative numerical
approach for studying the large scale, high frequency 
driven capacitively coupled discharges, which is able 
to describe the electromagnetic phenomena occurring
in such discharges. The approach is based on the Darwin 
approximation for solving the system of Maxwell's 
equations \cite{darwin1920}. The Darwin approximation reduces the original 
system of hyperbolic equations to a set of elliptic
equations by neglecting the transversal part of the
displacement current in Ampere's law, which is justified
by a scale analysis of the phenomena of interest.
Since the resulting elliptic equations do not support
electromagnetic radiation in vacuum, the corresponding
Courant criterion is removed and the simulation time step
can be chosen as in electrostatic simulations. The required
numerical scheme is a relatively simple and straightforward
modification of the well-established electrostatic algorithms.
Whereas numerical approaches based on the Darwin 
approximation have proven successfull in unbounded plasmas
(see, e.g., Ref. \cite{schmitz2005}), application of 
the Darwin approximation in plasmas having plasma-wall interfaces 
is more intricate. It is therefore important that the approach proposed in this
article avoids the computationally expensive iterative
schemes used in the previous implementations of the Darwin 
approximation-based aglorithms for bounded plasmas (see Refs. \cite{hewett1987,
hewett1994, gibbons1995,gibbons1997}).

We demonstrate that the Darwin approximation is in very
good agreement with the fully electromagnetic 
linear treatment of the problem described in
\cite{lieberman2002}. However, realistic plasma 
density profiles are far from the model used 
in the latter work, hence next we study
the standing wave and skin effect phenomena employing
a nonlinear fluid model based on the Darwin approximation
with more realistic plasma profiles treated in time domain.
Unlike some works (see, e.g., Ref. \cite{yang2010}), which consider
excitation of the electromagnetic fields in the discharge
coming from external sources, we assume the ``natural"
excitation of the electromagnetic fields in the CCP
discharge generated by the voltage/current that drives
the discharge. Obviously, such kind of excitation is
always present in the discharge. Also, our Darwin-fluid
model accounts for finite electron inertia effects which 
is necessary in plasmas with low collisions driven at high frequencies. 

The paper is organized as follows: In Section II
we briefly describe the assumptions of the Darwin
approximation and its applicability. In Section III 
we describe the choice of representation for the 
electromagnetic fields that is particularly convenient 
for the numerical implementation and a model geometry
used in the examples studied in this paper,
the boundary conditions are also given in this
section. In Section IV  we use a linear fluid model
solving for the electromagnetic fields in frequency
domain both with the Darwin approximation and the full system
of Maxwell's equations for two example cases with
pronounced standing wave and skin effects, respectively,
to corroborate that the Darwin approximation gives an accurate
description of the phenomena in question for the parameters
of interest. In Section IV we describe a nonlinear fluid
model accounting for the finite electron inertia effects
and also using the Darwin approximation to solve for the
electromagnetic fields. We demonstrate in a model simulation
the standing wave effect for a case with more realistic
plasma density profiles and sheath dynamics than in cases 
investigated in Section IV. Finally, we summarize the main
results of our study in Section VI.


\clearpage

\section{The Darwin approximation of the Boltzmann-Maxwell system}


In general, the electrodynamic behavior of any high
frequency discharge can be described by the kinetic equation, such as 
Boltzmann's equation, supplemented with the system of Maxwell's 
equations along with appropriate boundary conditions.
Reduced fluid models can be constructed from this general
system by taking moments of Boltzmann's equation and
postulating closure relations. The dynamics of the 
particle distribution function of charged species $s$ 
under influence of electromagnetic forces and
collisions is determined by Boltzmann's equation, 
\begin{gather}
\frac{\partial f_s}{\partial t} +
\vec v \cdot \nabla f_s + 
\frac{Z_s q_s}{m_s}\left(\vec E 
+\vec v\times\vec B\right)
\cdot \nabla_{\! v} f_s
= \left.\frac{\partial f_s}{\partial t}
\right|_{\text{col}}. \label{2-1}
\end{gather}
In turn, the electromagnetic field is described by
Maxwell's equations
\begin{gather}
\frac{1}{\mu_0}\nabla\times\vec B = 
\vec j + \epsilon_0 
\frac{\partial\vec E}{\partial t}, \label{2-2} \\
\epsilon_0\nabla\times\vec E =
- \frac{1}{c^2\mu_0}\frac{\partial \vec B}
{\partial t} \label{2-3} \\
\nabla\cdot\vec B =0,  \label{2-4} \\
\epsilon_0\nabla\cdot\vec E =\rho, \label{2-5}
\end{gather}
where $\rho = \sum_s Z_s q_s \int f_s  d^3 v$ and 
$\vec j = \sum_s Z_s q_s 
\int \vec v_s f_s d^3 v$.


Direct attempts at numerical solution of
\eqref{2-1}-\eqref{2-5} usually require very long 
simulation time. This is connected to the fact
that this system supports electromagnetic waves
in vacuum, which must be resolved in explicit algorithms
in order to avoid numerical instabilities. Another approach
is to use an implicit algorithm for the numerical solution, 
where unresolved electromagnetic modes of no interest are
damped numerically. However, such algorithms are cumbersome
to implement and often resort to artificial numerical
constructions (see, e.g., \cite{bowers2002}).

A conceptually different alternative is to remove
the stiffest time scale in electromagnetic simulation
by reduction of the original system of equations
using the assumption about the time and spatial scales
of the expected phenomena. Namely, we use the
fact that for the far most CCP discharges of interest 
the electrode size is small compared to the wavelength
of the electromagnetic wave in vacuum corresponding
to the highest frequency driving harmonics. 

Following the logic of \cite{raviart1996}, one can
recast \eqref{2-2} and \eqref{2-3} into a normalized
form where $\epsilon = L/c t$ with $L$ and $t$ the typical
spatial and time scale of the problem,
\begin{gather}
\nabla\times\vec B = \vec j + \epsilon 
\frac{\partial\vec E}{\partial t}, \label{2-8} \\
\nabla\times\vec E =- \epsilon \frac{\partial
\vec B}{\partial t}. \label{2-9}
\end{gather}
To the zeroth order in $\epsilon$ \eqref{2-9} yields 
$\nabla\times\vec{E}=0$, so that to this order the
electric field is purely longitudinal (irrotational);
we will denote this part of the electric field
$\vec{E}_L$. To the same order the displacement
current drops out of \eqref{2-8}. To describe the 
electromagnetic effects one needs to include the
next order in $\epsilon$, whereby  
\begin{gather}
\nabla\times\vec B = \vec j + \epsilon 
\frac{\partial{\vec E}_L}{\partial t}, \label{2-10} \\
\nabla\times{\vec E}_T =- \epsilon 
\frac{\partial \vec B}{\partial t}, \label{2-11}
\end{gather}
with ${\vec E}_T$ the transversal (solenoidal) part 
of the electric field, $\nabla\cdot{\vec E}_T=0$.
\eqref{2-10} and \eqref{2-11} constitute the Darwin
approximation.

One can see that this model eliminates electromagnetic
waves, but keeps an important part of the physics,
namely electromagnetic effects in the relatively low
frequency phenomena (when the corresponding wavelength
in vacuum is larger than the system size). Note that
the corresponding frequencies are called (very) high
frequencies in the standard terminology of the CCP
discharges. It is important to note that by dropping the 
transversal part of the displacement current the continuity
equation remains satisfied.

The relation of the Darwin model to the full system
of Maxwell's equations can also be illustrated
by the cold unmagnetized plasma dispersion relation in an unbounded plasma, 
\begin{gather}
c^2 k^2+\frac{i\omega  \omega_{\rm pe}^2}{i\omega+\nu}
-\omega^2 = 0.
\end{gather}
Fig. 1 displays the solutions $k=k(\omega)$ of this
relation for a typical parameter combination
($\omega_{\rm pe}= 2\pi \times 1\,{\rm GHz}$,
$\nu/\omega_{\rm pe}=0.01$). 
Three distinct regimes can be discerned. Slow phenomena
($\omega\le\nu$) are governed by the collisional skin effect
$k = (1-i)/\sqrt{2} \sqrt{\omega/\nu }\,
\lambda_{\rm scf}^{-1}$, with $\lambda_{\rm scf} = c/\omega_{pe}$
the collisionless skin-depth.
In the intermediate frequency range
($\nu\le\omega\le\omega_{\rm pe}$), the collisionless skin
effect prevails, $k= i \lambda_{\rm scf}^{-1}.$
Finally, for large frequencies ($\omega>\omega_{\rm pe}$)
this dispersion relation describes propagation of the
electromagnetic waves having
$k= \omega/c$.

In contrast, the dispersion relation of the Darwin model is
\begin{gather}
	c^2 k^2 +\frac{i 
	   \omega  \omega_{\rm pe}^2}{i \omega+\nu } =0.
\end{gather}
The Darwin model then covers the middle ground between the
electrostatic approximation and a fully electromagnetic
treatment. This example, however, serves only illustrative purpose
and is not general in stating that the Darwin approximation is
inapplicable for $\omega>\omega_{pe}$ as the dispersion relation
of waves in bounded plasmas can significanly differ from Eq. (11)
(as is the case in the present study, for example). The decisive conclusion regarding applicability
of the Darwin approximation for description of a phenomenon in 
question should be made on the basis of the smallnes of the $\epsilon$
parameter defined above.

Historically, Darwin's original derivation \cite{darwin1920} was designed to
describe a set of charged particles in free space with
velocities small compared to the speed of light. He started
from the Lagrangian description of the $N$-body problem
with the interaction incorporated via the retarded
Lienard-Wichert potentials. A formal expansion in the
smallness parameter $v/c$, carried up to the second order,
resulted in a simplified Lagrangian with instantaneous 
(not retarded) potentials. In the same expansion order,
relativistic corrections to the equations of motion appeared.
For a recent analysis of that approach, see \cite{krause2007}.
  

\clearpage

\section{Problem Setup}


For the sake of concreteness, we will consider the 
same problem setup for studying the standing wave and
skin effect phenomena in CCP discharges as was
suggested by Lieberman et al. \cite{lieberman2002}, i.e.,
a  CCP discharge having cylindrical geometry symmetric
in the azimuthal direction bounded by a dielectric 
sidewall (see Fig. 2). 

As all the numerical models
described in this article will use the same field equations 
stemming from the Darwin approximation of Maxwell's equation,
we will describe them in this section. We saw in the previous section that the electric field
in the Darwin approximation is separated into two parts,
a longitudinal part, which is curl-free, and a transversal
part, which is divergence-free. Hence, we choose to represent
the electric field in the following way,
\begin{gather}
\vec{E} = -\nabla\phi + \nabla\times
{\vec A_T}, \label{3-1}
\end{gather}
where the first (second) term expresses the longitudinal
(transversal) part of the electric field. The longitudinal
potential $\phi$ is governed by Poisson's equation,
\begin{gather}
\nabla^2 \phi = -\frac{\rho}{\epsilon_0}, \label{3-2}
\end{gather}
with $\rho$ the charge density. To find an equation
for the transversal vector-potential $\vec A_T$ we first
take the curl of Ampere's law in the Darwin approximation,
which results in
\begin{gather}
\nabla^2  \frac{\partial \vec{B}}{\partial t} = -\mu_0 \nabla 
\times \frac{\partial \vec j}{\partial t}, \label{3-3}
\end{gather}
where we used \eqref{2-4}. Then, one can exploit
Faraday's law using the chosen  representation for the 
transversal field,
\begin{gather}
\nabla\times \nabla \times \vec A_T = -\frac{\partial \vec B}{\partial t}. 
\label{3-4}
\end{gather}

Following \cite{lieberman2002}, we consider only TM modes
with $\vec{E}  = \vec e_r E_r + \vec e_z E_z$ and 
$\vec B = \vec e_\theta B_\theta$, and $\vec A_T= 
\vec e_\theta A_{T\theta}$, so that $\nabla\cdot \vec B=0$
and $\nabla\cdot \vec A_T = 0$  hold automatically due
to the symmetry of the problem.

In all the following models we study the electromagnetic
response to the ``natural excitation'' of a CCP discharge, 
prescribing a harmonically varying constant magnetic field
at the sidewall, which is in agreement with the previously
made assumption that all the RF current, which drives the
discharge, flows through it, and no external excitation source 
is involved. Correspondingly, boundary conditions for 
$B_\theta$ is $\partial_z B_\theta = 0$ at 
the electrodes ($z=-l$ and $z=l$), which is needed in order
for the tangential electric field to vanish at the electrodes,
and reads $B_\theta=\mu_0 I/(2\pi R)$ at the sidewall 
($r=R$)  (provided no RF current leaves the 
discharge through the dielectric sidewall).
Boundary conditions for $\partial_t B_\theta$ are 
obtained by taking time derivative of the boundary conditions 
for $B_\theta$. Further, boundary conditions for 
$A_{T\theta}$ are $\partial_z A_{T\theta} = 0$ 
at the electrodes, so that the tangential component of
the transversal electric field vanishes, and $A_{T\theta}=0$
at the sidewall. The latter boundary
condition for the $A_{T\theta}$ at the sidewall can be 
obtained as follows. From the condition that the radial component of
the transversal electric field vanishes at the dielectric sidewall one
can deduce that $\partial_z A_{T\theta}=0$ at $r=R$. 
Integrating this equation with respect to $z$ at $r=R$, matching
the boundary conditions for $A_{T\theta}$ at the electrodes and
setting the common constant to zero, one obtains the boundary
condition sought. 

As the boundary conditions for the longitudinal potential
will be slightly different in the two models studied in this 
work, we will specify the boundary conditions for each
particular model separately. 


\clearpage 

\section{Linear model problem with fixed sheath treated in
frequency domain}


In this section we will demonstrate that the Darwin 
approximation is a valid approach for investigation of 
the electromagnetic phenomena in CCPs by comparing the 
full electromagnetic solution to solution obtained with 
the Darwin approximation for the problem investigated 
by Lieberman et al. \cite{lieberman2002}.  The model plasma studied there has uniform
ion density and a stepwise density for the electron 
component, such that electrons have the same density as ions 
in the bulk and zero density in the sheath, the sheath boundaries 
are considered to be stationary. The plasma is modeled as a
dielectric, change of the electric field in the sheaths due 
to motion of the bulk electron plasma component is modeled
through the surface charge created at the interfaces between 
the bulk plasma and the sheath regions. This approach is valid for 
small oscillations of the bulk plasma (see \cite{bowers2002}), 
a more general case of moving sheath boundary is considered
in the next section. 

Then, using the representation of the electric field given
in \eqref{3-1} (which holds in a general case as well), one can
obtain the following system of equations,
\begin{gather}
\nabla\times\frac{1}{\epsilon_L} \nabla\times\vec B 
= \frac{i\omega}{c^2}\nabla\times\frac
{\epsilon_T}{\epsilon_L}\nabla \times 
\vec A_T \label{4-1} \\
\nabla\times \nabla\times\vec A_T = 
-i\omega\vec{B} \label{4-2}
\end{gather}
provided all quantities are proportional 
to $e^{i\omega t}$. In \eqref{4-1} and \eqref{4-2}, 
$\epsilon_L = 1$ in the sheath and $\epsilon_L=1 
- \omega_{pe}^2/\omega(\omega-i\nu)$ in the bulk.
This system of equations describes both fully EM case 
(if $\epsilon_T = \epsilon_L$) and the Darwin approximation 
(if $\epsilon_T = 0$ in the sheath and 
$\epsilon_T=-\omega_{pe}^2/\omega(\omega-i\nu)$ in the bulk; 
this expression can be easily obtained in the usual way by
using the equation of motion and Ampere's law in the Darwin approximation).

Once one obtains $\vec A_T$ and $\vec B$ from solution 
of \eqref{4-1} and \eqref{4-2}, the transversal electric 
field can be obtained from $\vec E_T = \nabla\times \vec A_T$ and 
the longitudinal electric field from Ampere's law, 
which provides $ \vec E_L = -ic^2/\omega\epsilon_L \nabla
\times\vec B - \epsilon_T/\epsilon_L \nabla\times \vec A_T$.

Results in Figs. 3 and 4 show the power deposition profiles for two selected
cases with similar parameters to the corresponding cases in 
\cite{lieberman2002}. In all cases $d=2\,{\rm cm}$ $s=0.4\,{\rm cm}$, and $\nu=10^7\,{\rm s}^{-1}$.
The radial profiles shown in Figs. 3 and 4 are normalized values of $P_{cap}=\int^d_0 |E_z|^2 dz$ (with $E_z$
being predominantly capacitively generated electric field)
and  $P_{ind}=\int^d_0 |E_r|^2 dz$ \cite{lieberman2002} (with $E_r$ being predominantly induced 
electric field), calculated with the full system of Maxwell's equations and 
the Darwin approximation. Not only do both models demonstrate
same behavior qualitatively, but also quantitative difference 
lies within a numerical error for the parameters chosen. 
Consequently, from this moment on we will suppose that the Darwin 
approximation is a valid approach for description of the 
standing wave and skin effects.


\clearpage

\section{Nonlinear model problem with moving sheath treated in time domain}


Although the model problem used \cite{lieberman2002}
and in the previous section gives a certain insight into 
the standing wave and skin effects, it is still quite far
from realistic CCP plasmas. The assumptions of 
stationary sheath boundary and stepwise electron density
profile are the main drawbacks of the model. Most of the 
previous analytical and numerical treatments of the standing wave and skin effects in 
the literature also assumed stationary sheath boundary
with the electrostatic sheath model (e.g., \cite{lee2008}), 
which makes the corresponding numerical algorithm faster, 
but strips the model of adequate description of possible 
nonlinear phenomena. In  \cite{zhang2010} the
authors studied the problem entirely in the time domain, 
however,  the commonly adopted drift-diffusion approximation 
also used in that work is clearly not applicable to the 
usual range of parameters in low-pressure CCP discharges, 
when the frequency of electron collisional momentum transfer is
smaller than the driving frequency. In particular, the 
highly collisional regime studied in \cite{zhang2010} 
renders direct comparison with results of \cite{lieberman2002}
impossible, as the latter work was made under the 
low-collisionality ansatz.

To fully resolve the nonlinear discharge dynamics we 
propose another model, where the sheath boundary is 
allowed to move, the plasma density profiles have a more 
realistic shape,  and the effects of finite electron 
mass are retained, which makes the model capable of 
more general treatment of the electromagnetic phenomena under question 
in low-pressure CCP discharges than the model considered in 
the previous section. The electromagnetic 
fields in the sheath region are calculated self-consistently 
with the plasma dynamics and are treated in the same way 
as the rest of the electromagnetic fields in the 
entire discharge. To keep the model simple yet 
comprising all the effects of interest, we adopt a fluid 
description for the electron plasma component, whereas 
ions are assumed to be immobile. To make the model
realistic, we adopt the axial ion density profile from a 2D electrostatic
PIC/MCC simulation (with the same geometry and plasma
parameters as those of the fluid simulation studied here), 
taken at the half-radius radial position (since
the density profiles in the electrostatic case are essentially
uniform, we feel that such an approach is justified).
In this way we choose 
to focus on the coupled dynamics of electrons and the 
electromagnetic fields in the entire discharge, omitting 
the issue of discharge sustainment. 
Analogously, we substitute the plasma energy balance
equation with the assumption of electron isothermality. 
The model geometry is the same as considered in the 
previous section.

The evolution of electrons is governed by the momentum equation,
\begin{gather}
\frac{\partial \vec v_e}{\partial t} + 
(\vec v_e \cdot \nabla)\vec v_e = 
-\frac{T_e}{m_e}\frac{\nabla n_e}{n_e}
-\frac{e}{m_e}\left(-\nabla \phi +\nabla\times\vec 
A_T + \vec v_e\times \vec B \right) 
- \nu_m \vec v_e, \label{5-1}
\end{gather} 
and the continuity equation,
\begin{gather}
\frac{\partial n_e}{\partial t} 
= -\nabla\cdot (n_e\vec v_e). \label{5-2}
\end{gather}
The boundary conditions for \eqref{5-1} is that
the radial component of the electron  velocity, $v_{e,r}$, 
vanishes at the sidewall, which comes from the assumption 
that all the electrons are reflected back into the discharge
there, and that the axial component of the 
electron velocity at the driven (grounded) electrodes is 
$v_{e,z} = \pm 1/4\sqrt
{8k_BT_e/\pi m_e}$, respectively, which corresponds 
to the kinetically limited flux. 
The boundary conditions for \eqref{5-2} are, in 
accordance with the boundary conditions 
for \eqref{5-1}, that the radial flux to the sidewall 
vanishes, and the axial flux to the electrodes is kinetically
limited, so that $n_e v_{e,z} = \pm n_e/4\sqrt{8k_BT_e/\pi m_e}$ 
at the driven (grounded) electrodes, respectively. 
We take the initial ion density profile to be of 
the Gaussian form and the initial electron 
density profile to be of the same form as the
ion density profile. The numerical discharge simulation 
gradually develops electron-depleted 
sheath regions (see Fig.5). 

The fields are calculated self-consistently with evolution 
of the electron plasma component using  \eqref{3-2} to 
\eqref{3-4}). One can distinguish two different sets of 
boundary conditions depending on the way the 
discharge is driven, either with a voltage source or 
a current source. Whereas boundary conditions for 
\eqref{3-4} are same for both cases (see the discussion 
at the end of section III), the boundary conditions 
for \eqref{3-2} and \eqref{3-3} are different for each case.

\subsection{Voltage-driven discharge}

In this case the potential at the driven electrode 
is prescribed, $\phi(z=-l) = V(t)$. This  gives 
Dirichlet boundary condition at the driven electrode,
the boundary conditions at the sidewall and the 
grounded electrode being same in both cases 
(Neumann boundary condition at the sidewall, 
$\partial_r \phi = 0$ and at the 
grounded electrode Dirichlet boundary condition 
$\phi = 0$. (See also the discussion in Section III). 

The time derivative of the total axial current 
through the discharge, which enters the sidewall 
boundary condition for \eqref{3-3}, must be 
calculated under the circumstances
by integrating the plasma current density time 
derivative, $-e \partial_t (n_e v_{e,z})$  
using \eqref{5-1} and \eqref{5-2} over the surface
of any discharge cross-section where the plasma 
current is dominant, for example, in the midplane.

\subsection{Current-driven discharge}

The potential at the driven electrode is calculate 
in this case from Kirchhoff's law of 
current continuity at the driven electrode, where the
sum of the electron current and the displacement 
current in the sheath must match the 
current supplied by the source. 
We seek solution to $\phi$ in the form $\phi = 
(\phi_0+\tilde\phi(t))\Phi_{\rm v} + \Phi_
\rho$, where $\nabla^2 \Phi_{\rm v} = 0$, where 
$\phi_0$ is the stationary, and $\tilde\phi$ 
the harmonically changing parts of the 
potential at the driven electrode, respectively. The boundary conditions
are: $\Phi_{\rm v}=1$ at the driven electrode,
$\Phi_{\rm v}=0$ at the grounded electrode, and 
$\partial_r \Phi_{\rm v} = 0$ at 
the sidewall;  $\Phi_\rho$ satisfies $\nabla^2 \Phi_\rho = -\rho/\epsilon_0$ 
with  boundary conditions  $\Phi_
\rho=0$ at both electrodes, and $\partial_r \Phi_\rho = 0$ 
at the sidewall. 
From the current balance at the driven electrode one has
$-2\pi\int_0^R  (e n_e v_e + 
\epsilon_0 \partial_t\partial_z \phi)rdr = I(t)$.  
Substituing the chosen form for $\phi$ into this equation, 
one obtains an equation for the potential at the driven 
electrode
\begin{equation}
\frac{\partial \tilde\phi}{\partial t} 
= - \frac{I(t) + 2\pi\displaystyle\int_0^R \left( e 
n_e v_{e,z} + \epsilon_0 \partial_t\partial_z 
\Phi_\rho\right)rdr}{2\pi
\epsilon_0\displaystyle\int_0^R  \left(\partial_z\Phi_{\rm v}\right) rdr}  
\end{equation}

The boundary condition for \eqref{3-3} at the sidewall 
is particularly simple in the case 
of current-driven discharge, as one can directly use the
value of the time derivative of the total current 
supplied by the source.

Finally, we want to briefly discuss some details 
of numerical implementation of \eqref{3-2}), \eqref{3-3},
\eqref{3-4}, \eqref{5-1}, and \eqref{5-2}. We use a 
leapfrog-like scheme to integrate \eqref{5-1} and \eqref{5-2}
in time, for which we use grids staggered in time for 
velocity and density, the density and field values being 
taken at integral time levels and velocity at the half-
time levels. Then, the order of the whole numerical
scheme is the following:

\begin{enumerate} 
\item \eqref{5-1} is solved treating the nonlinear 
second term on the left-hand-side 
semi-implicitly thereby linearizing this term 
with respect to the 
density taken at the new time level. The velocity 
on the right hand side is time-centralized by 
taking the average between the new and old time levels, so that it is also 
semi-implicit. It is worth mentioning that during 
the stage of the sheath formation an artificial viscosity 
term is helpful in getting rid of numerical instabilities 
arising at the plasma-sheath interface. Once smooth 
sheaths are formed, the viscosity term 
can be switched off. 
\item The density on the right-hand side of \eqref{5-2}
is treated semi-implicitly and is time-centralized using 
the same technique as in the previous procedure.
\item \eqref{3-2} is solved with the boundary conditions 
depending on the way the discharge is driven according 
to the techniques described above.
\item The source term in \eqref{3-3} should be calculated 
using the expression $\partial_t \vec j_e 
= -e (\vec v_e \partial_t n_e + n_e \partial_t \vec v_e)$ with $
\partial_t \vec v_e$ and $\partial_t n_e$ calculated with
help of \eqref{5-1} and \eqref{5-2}. An attempt at direct 
evaluation of $\partial_t \vec j_e$ will lead to numerical 
instability because numerical discretization of a time derivative
introduces numerical dispersion and finite field propagation 
time, whereas conceptually the Darwin approximation expects instantaneous 
field change in the whole domain (see, e.g., \cite{nielson1976}).
When calculating the source term for \eqref{3-3}, it is also 
useful to write the term $\mu_0 \nabla\times e^2n_e/m_e 
\nabla\times\vec A_T $ as $-\lambda_{\rm scf}^2\partial_t\vec B +  
\mu_0 e^2n_e/m_e \nabla n_e \times \nabla \times \vec 
A_T$, where $\lambda_{\rm scf} = c/\omega_{pe}$ is the collisionless 
skin depth and we used \eqref{3-4}.  The first term in 
this expression describes the collisionless skin effect and can 
be transferred to the left hand side to be treated implicitly
for better numerical stability.
\item Finally, \eqref{3-4} is solved using the boundary conditions 
discussed in Section III.
\end{enumerate}

One can see that because of the chosen representation 
for the transversal field as $\vec E_T = \nabla  \times \vec A_T$ 
the cumbersome and computationally expensive iterative
scheme needed for making sure that the calculated $\vec E_T$ 
is divergence-free used in the previous implementations 
of the Darwin codes (see, e.g., 
\cite{hewett1987, hewett1994, gibbons1995, gibbons1997})
is avoided. This is because $\vec E_T$ calculated by 
taking the curl of $\vec A_T$ guarantees that $\vec E_T$
is divergence-free automatically. To account for the 
electromagnetic effects in 2D geometry of the problem
we described, one needs to solve only two additional elliptic 
equations using the same field solver that is used for 
solving Poisson's equation. It is a small price for
removing the very expensive Courant criterion
connected with propagation of the light wave, though.

In Fig. 6 we show the radial profile of relative power
deposition due to the Ohmic heating by the axial electric field
resulting from a 
current-driven discharge simulation showing a significant
standing wave effect for a case with relatively low
collisionality ($\nu_m=1\times 10^7$ s$^{-1}$). 
Parameters are similar to the parameters for the
case shown on the top part of Fig. 3. The inductive power 
deposition is negligible and is not shown. 
For comparison we show results of the linear problem of the Section IV  with the stepwise electron density profile
and equivalent sheath width $s$ calculated from the equation
$\int_s^{l/2} dz (\overline{n}_i(z)-\overline{n}_e(z)) = \int_0^s dz \overline{n}_e(z)$  (see \cite{brinkmann2009}) with
the period averaged densities shown in Fig. 5.
The non-uniformity of the
radial power deposition profile  is slightly more pronounced
in the case of the nonlinear fluid model of the present Section with the more realistic electron axial density profile,
but in general
both models are in good agreement. Note that the 
radial profile of relative power deposition obtained with
a purely electrostatic simulation would be uniform. As one can see, the 
electromagnetic component is an essential part of the standing wave 
effect for the parameters in question, despite the fact that
surface waves propagating in sheath and generating standing
waves may also exist in the electrostatic description (see, 
e.g., \cite{cooperberg1998}). 

The evidence of the model nonlinearity can be seen on the Fourier
spectrum of the time-dependent voltage drop across the current-driven discharge (Fig. 7).
One can clearly see that apart from the fundamental harmonic, 
the spectrum exhibits response at the third harmonic also. The amplitudes
of other harmonics are noticeably supressed, which might be attributed to
the fact that the fluid models tend to over-relax the arsiing density gradients
leading to damping of high harmonics. 


\clearpage

\section{Conclusions and outlook}


In this paper we have demonstrated that the Darwin 
approximation is adequate for description 
of the standing wave and skin effects present 
in the high frequency CCP discharges by 
comparing relevant electromagnetic fields calculated 
using the approximation and solution of 
the full set of Maxwell's equations for two 
different typical cases exhibiting such effects 
in a model problem. We considered the ``natural'' 
excitation of such effects coming from the 
current driving the discharge needed for its 
very existence. The Darwin approximation 
reduces originally hyperbolic Maxwell's equations, 
by dropping the transversal current in 
Ampere's law, to a set of elliptic equations which 
do not support light waves propagating 
in vacuum. This has very significant consequence 
for the numerical algorithms based on that 
approach that the corresponding Courant criterion 
is also removed, enabling same time step 
size as in electrostatic algorithms.  We have
shown that the Darwin approximation can be 
straightforwardly implemented in 
fluid models, requiring only a slight 
modification of a possibly existing electrostatic code 
with the eventual great computational pay-off.  
The choice of the representation for the 
transversal field proposed by us helps to 
avoid the computationally demanding iterative 
schemes needed for its divergence cleaning 
used in other implementations.
Using a reduced nonlinear fluid code 
self-consistently evolving plasma and electromagnetic 
fields in time domain, on the example of a case 
exhibiting standing wave effect we have 
shown that the electromagnetic effects persist 
for more realistic profiles and sheath dynamics.

The proper description of low-pressure CCP 
discharges ultimately requires a kinetic tool 
as kinetic effects are essential in such 
discharges. The use of the Darwin 
approximation is, of course, not limited to 
fluid models. The sheath dynamics, the heating 
mechanisms other than the Ohmic heating, the 
nonlocal transport phenomena all demand 
kinetic treatment. The Darwin numerical algorithm 
for the solution of the electromagnetic fields 
designed in this work can be also used in
a Particle-in-Cell code. The reduced fluid 
model developed in this work can, in turn,
serve as benchmark for such a code. 


\clearpage

\section*{Acknowledgments}

The authors gratefully acknowledge fruitful discussions with Prof. R. Grauer (Ruhr University Bochum) and
support by the Deutsche
Forschungsgemeinschaft via the Sonderforschungsbereich TRR 87.


\clearpage

\begin{figure}[t]
\centering
\includegraphics[width=15cm]{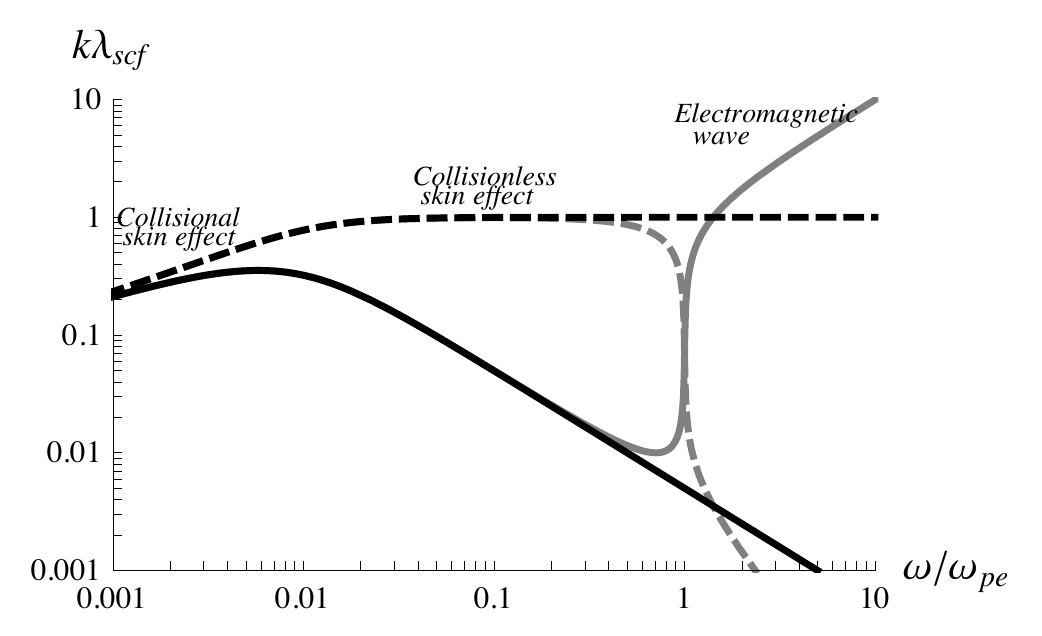}
\caption{The dispersion relation obtained with Darwin approximation closely follows the full electromagnetic 
dispersion relation as long as $\omega<\omega_{pe}$. Solid curves denote the corresponding real and
dashed ones imagn }
\end{figure}

\clearpage

\begin{figure}[t]
\centering
\includegraphics[width=15cm]{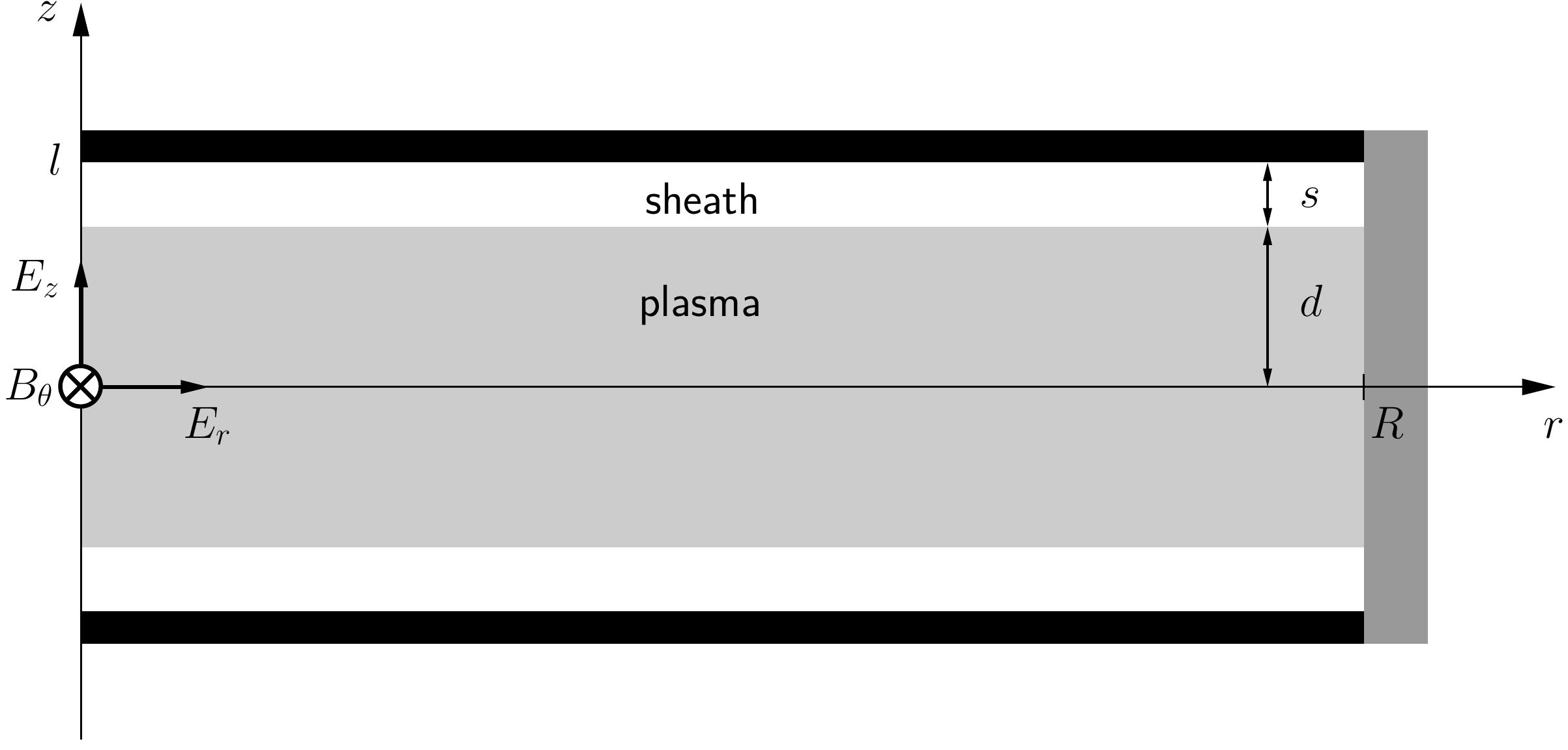}
\caption{Sketch of the problem geometry studied in both fluid models. The cylindrical CCP discharge under scrutiny has
symmetric electrodes, a sidewall made of dielectric and a plasma. The boundary at the sheath-plasma
interface is stationary in the model described in Section IV, whereas in the model of Section V it is allowed to 
move under the influence of electromagnetic fields. The electromagnetic fields under interest consist of
the TM modes in such a plasma-filled cavity and have nonzero $E_z, E_r$ components for the electric, and $B_\theta$ component for
the magnetic fields, respectively. }
\end{figure}

\clearpage

\begin{figure}[t]
\centering
\includegraphics[width=15cm]{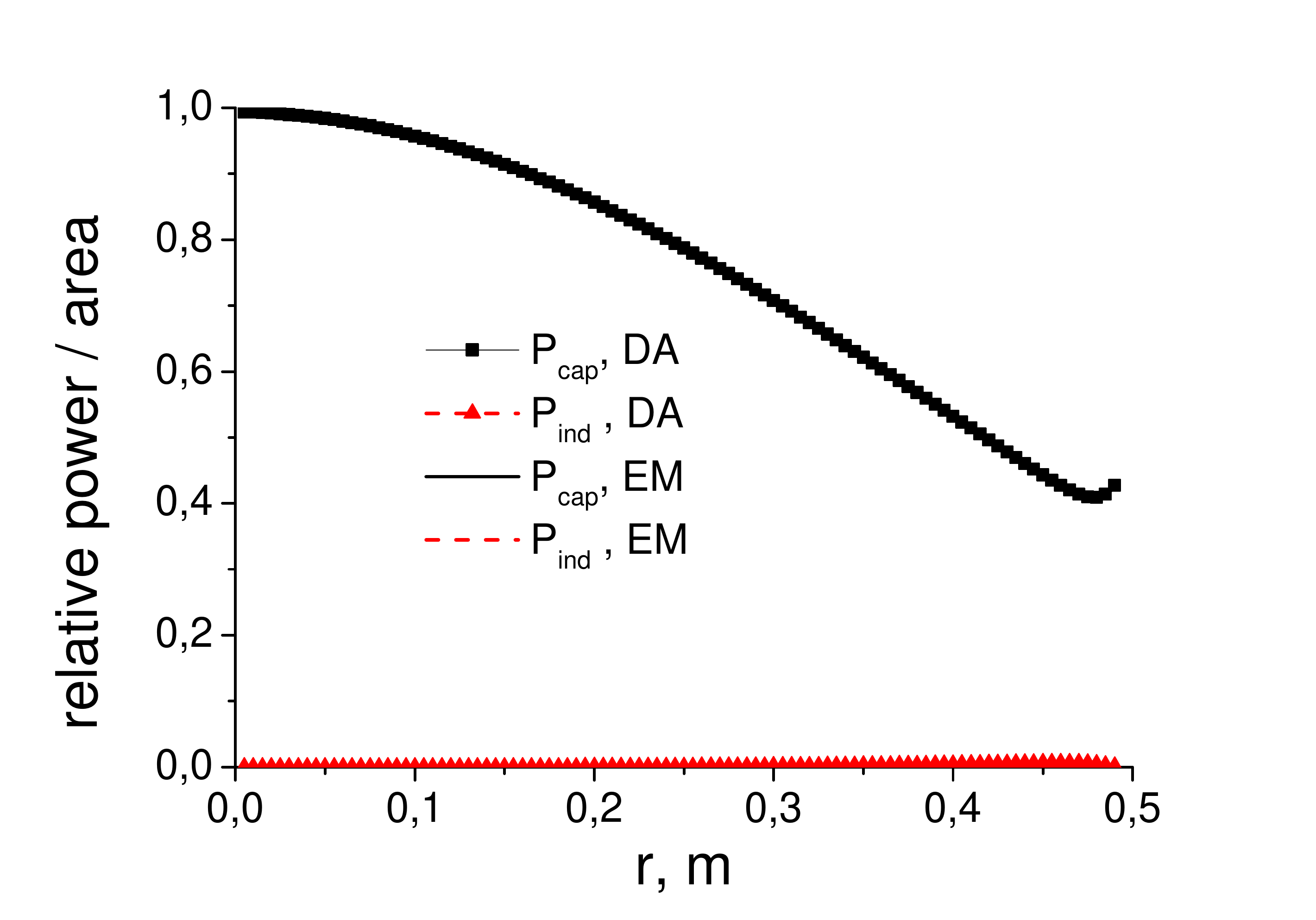}
\caption{Comparison of solutions obtained with Darwin approximation (DA) and full system of
Maxwell's equations (EM). The first case, which exhibits a strong standing wave effect, has
$f=40,7\,{\rm MHz}$ and $n=10^{15} \, {\rm m}^{-3}$. Here $P_{cap}=\int^d_0 |E_z|^2 dz$
and  $P_{ind}=\int^d_0 |E_r|^2 dz$, see also text.
}
\end{figure}

\clearpage

\begin{figure}[t]
\centering
\includegraphics[width=15cm]{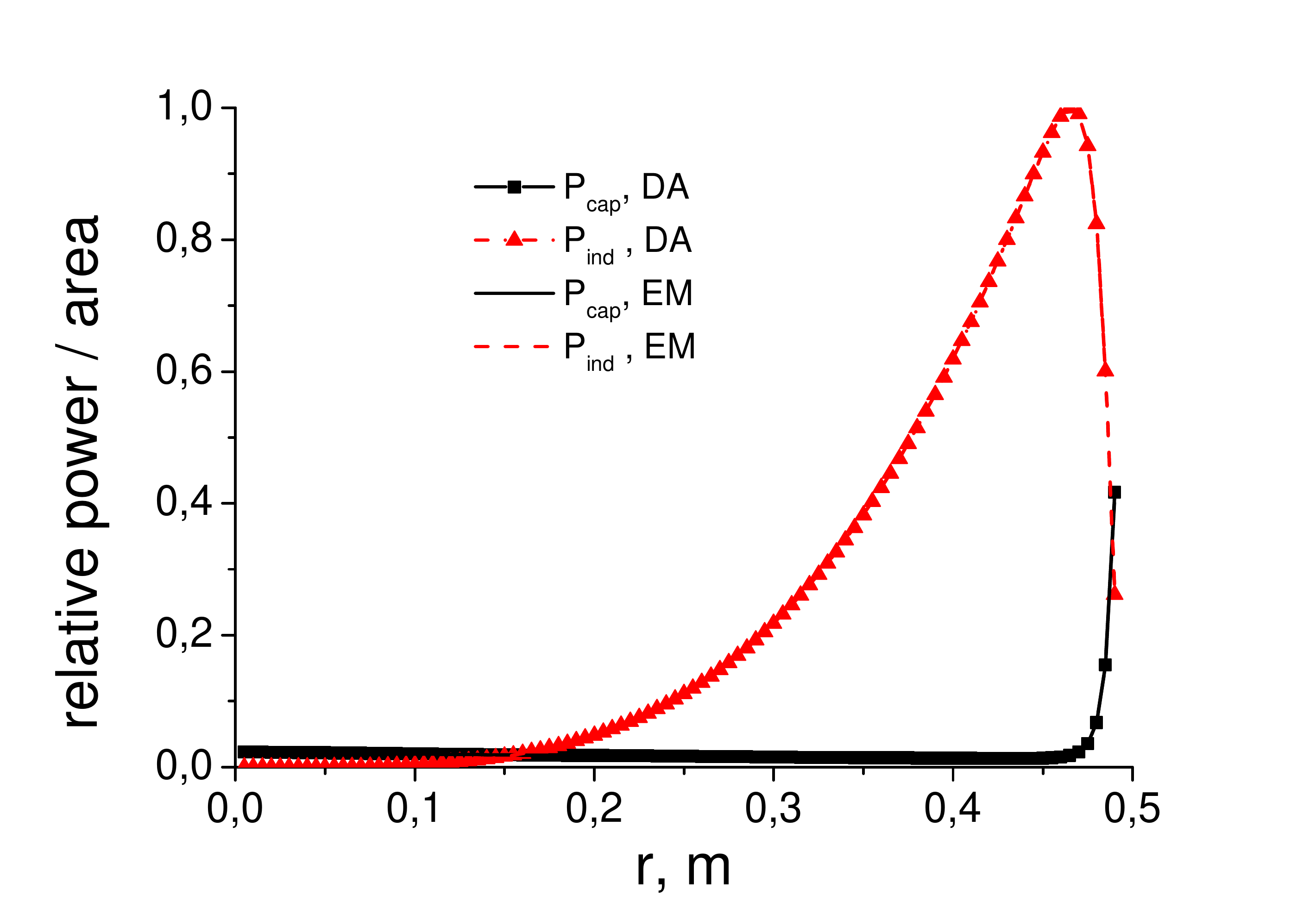}
\caption{The second case  shows a strong skin effect and
has $f=13.56\,{\rm MHz}$ and $n=10^{17} \, {\rm m}^{-3}$.}
\end{figure}

\clearpage

\begin{figure}[t]
\centering
\includegraphics[width=15cm]{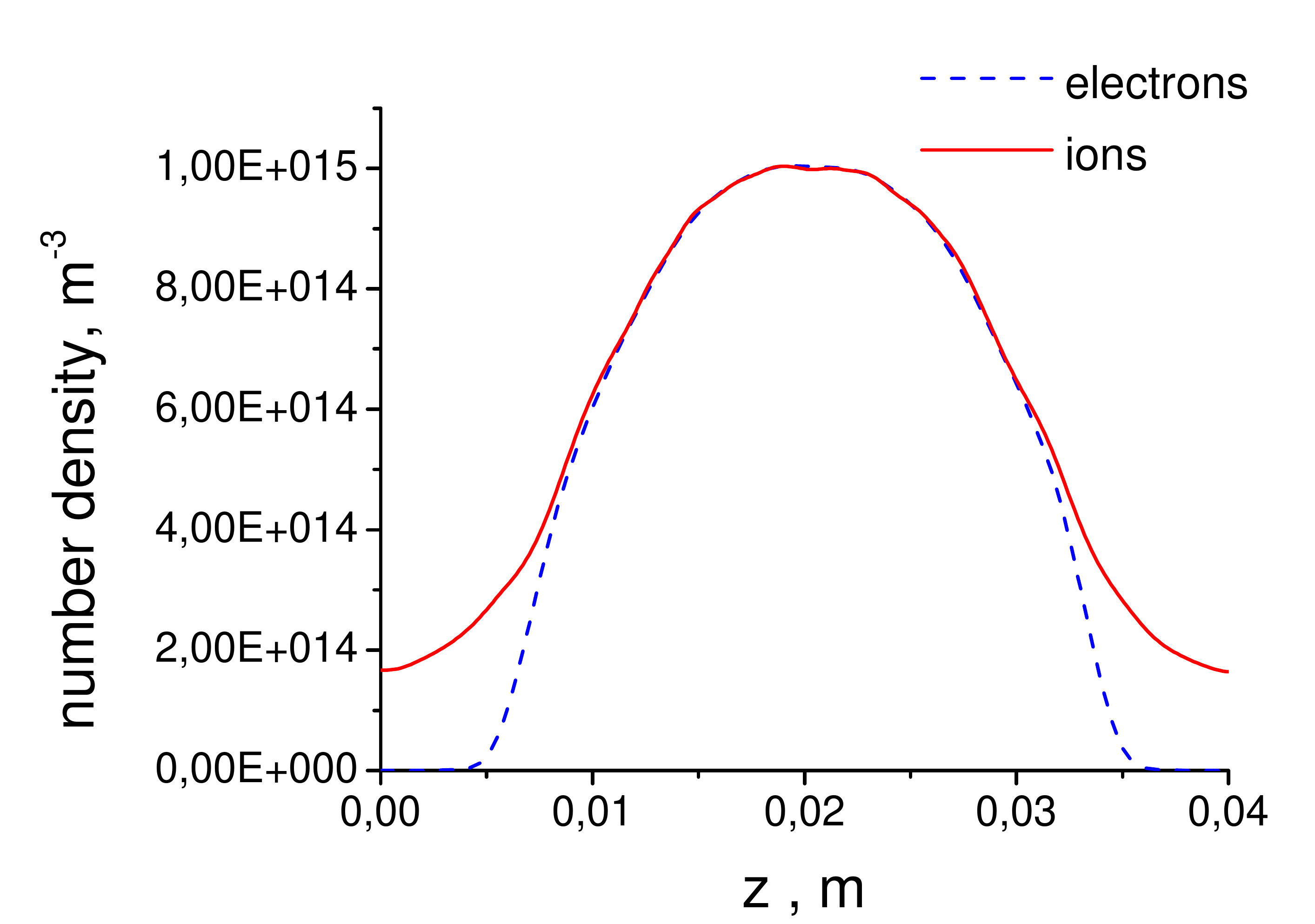}
\caption{In the model nonlinear simulations the ion density profile is taken from an externally conducted electrostatic PIC/MCC
simulation and is held fixed, whereas mobile electrons adjust themselves to create the electron-depleted sheath regions 
close to the electrodes. Shown is the electron density averaged over an RF period versus the ion density profile.
}
\end{figure}

\clearpage

\begin{figure}[t]
\centering
\includegraphics[width=15cm]{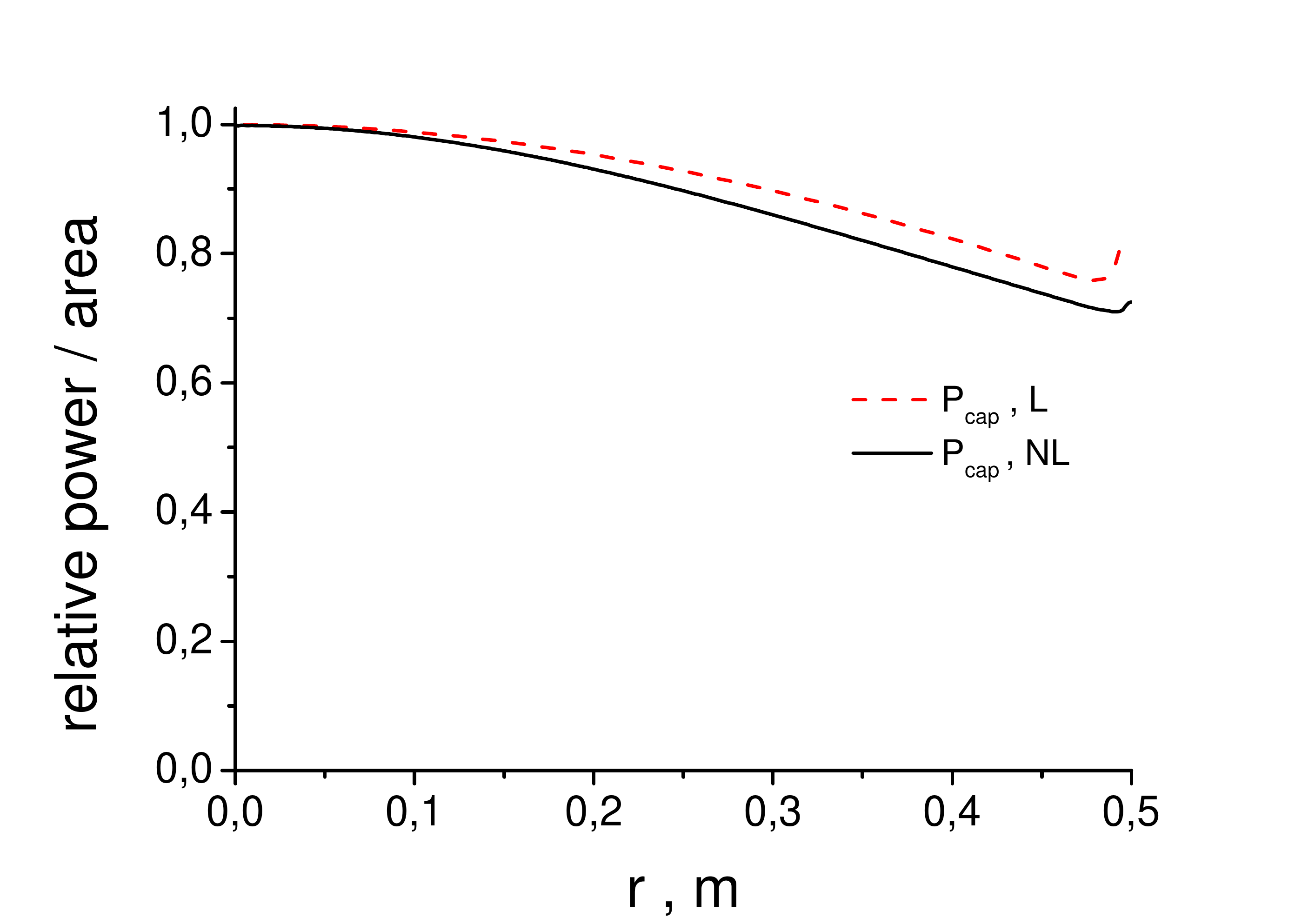}
\caption{Radial profile of the relative power deposition (Ohmic heating) performed by the axial electric field calculated from
the nonlinear problem (black solid curve) versus results of the linear problem (red dashed curve) with the equivalent sheath width. The case shown has $f=40.7\, 
{\rm MHz}$ and $n_{e,max}=1\times 10^{15} {\rm m}^{-3}$} 
\end{figure}

\clearpage

\begin{figure}[t]
\centering
\includegraphics[width=15cm]{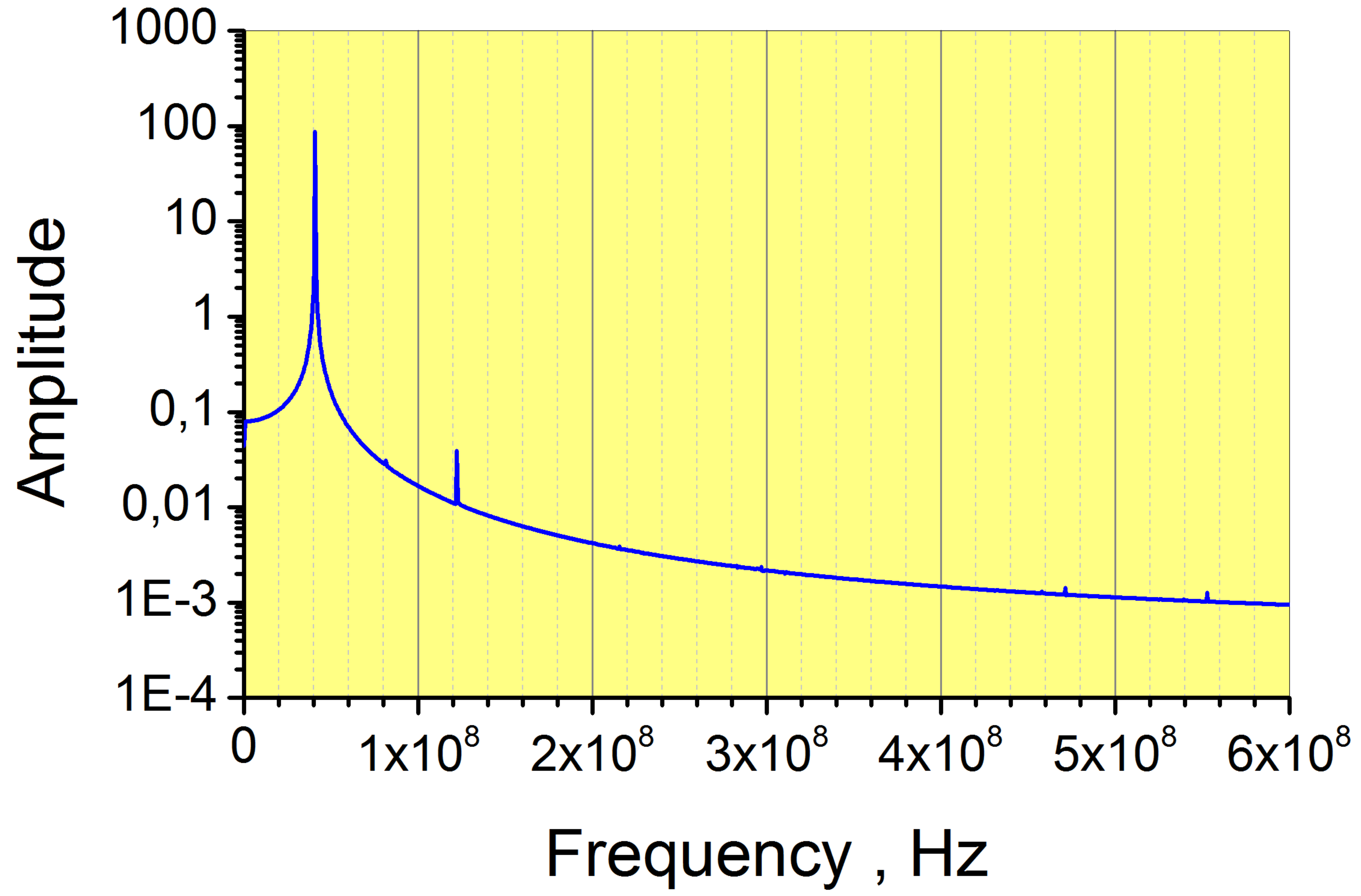}
\caption{Fourier frequency spectrum of the voltage drop time-dependent signal for the same current-driven discharge as shown in Fig. 6} 
\end{figure}

\end{document}